\documentclass[12pt]{article}
\usepackage{amsmath}
\usepackage{latexsym}
\usepackage{bbm}

\def\a{\alpha}

\def\b{\beta}
\def\g{\gamma}

\def\d{\delta}
\def\ep{\epsilon}
\def\e{\varepsilon}

\def\m{\mu}
\def\n{\nu}

\def\r{\rho}

\def\om{\omega}

\def\G{\Gamma}

\def\pa{\partial}

\def\ov{\overline}

\def\half{\frac{1}{2}}

\def\half{\frac{1}{2}}

\def\and{{\rm and}}

\begin{document}
\vspace*{-.6in} \thispagestyle{empty}
\begin{flushright}
CALT-68-2671
\end{flushright}
\baselineskip = 18pt

\vspace{1.5in} {\Large
\begin{center}
${\cal N} =8$ Superconformal Chern--Simons Theories\end{center}}

\begin{center}
Miguel A. Bandres, Arthur E. Lipstein and John H. Schwarz
\\
\emph{California Institute of Technology\\ Pasadena, CA  91125, USA}
\end{center}
\vspace{1in}

\begin{center}
\textbf{Abstract}
\end{center}
\begin{quotation}
\noindent A Lagrangian description of a maximally supersymmetric
conformal field theory in three dimensions was constructed recently
by Bagger and Lambert (BL). The BL theory has $SO(4)$ gauge symmetry and
contains scalar and spinor fields that transform as 4-vectors. We
verify that this theory has $OSp(8|4)$ superconformal symmetry and
that it is parity conserving despite the fact that it contains a
Chern--Simons term. We describe several unsuccessful attempts to construct
theories of this type for other gauge groups and representations. This
experience leads us to conjecture the uniqueness of the BL theory.
Given its large symmetry, we expect this theory to play a
significant role in the future development of string theory and M-theory.
\end{quotation}

\newpage

\pagenumbering{arabic}

\section{Introduction}

Following earlier studies of coincident M2-brane systems
\cite{Bagger:2006sk}, Bagger and Lambert (BL)
\cite{Bagger:2007jr,Bagger:2007vi} have constructed an explicit action
for a new maximally supersymmetric superconformal Chern--Simons theory in three
dimensions. The motivation for their work, like that in
\cite{Schwarz:2004yj}, is to construct the superconformal theories
that are dual to $AdS_4 \times S^7$ solutions of M-theory. Such
theories, which are associated to coincident M2-branes, should be
maximally supersymmetric, which in three dimensions means that they
have ${\cal N} =8$ supersymmetry. More precisely, the superconformal
symmetry group should be $OSp(8|4)$, which is also the symmetry of the
M-theory solution. It is not obvious that a classical action
describing the conformal field theory that is dual to the M-theory
solution needs to exist. In fact, there are good reasons to be
skeptical: These field theories can be defined as the infrared
conformal fixed points of nonconformal $SU(N)$ ${\cal N} =8$ Yang--Mills theories,
but there is no guarantee that any of these fixed points has a dual
Lagrangian description.

Ref.~\cite{Schwarz:2004yj} attempted to construct three-dimensional
theories with $OSp(8|4)$ superconformal symmetry and $SU(N)$ gauge
symmetry using scalar and spinor matter fields in the adjoint
representation of the gauge group. These would be analogous to ${\cal N} =4$
$SU(N)$ gauge theory in four dimensions, with one crucial difference. The
$F^2$ gauge field kinetic term has the wrong dimension for a
conformal theory in three dimensions. Also, it would give
propagating degrees of freedom, which are not desired. To address
both of these issues, \cite{Schwarz:2004yj} proposed using a Chern--Simons
term for the gauge fields instead of an $F^2$ term. The
conclusion reached in \cite{Schwarz:2004yj} was that such an action,
with ${\cal N} =8$ supersymmetry, does not exist. This was
consistent with the widely held belief (at the time) that
supersymmetric Chern--Simons theories in three dimensions only exist
for ${\cal N} \le 3$.\footnote{Theories of this type with ${\cal N} =2$
supersymmetry were first constructed by Ivanov \cite{Ivanov:1991fn}
and by Gates and Nishino \cite{Gates:1991qn}. For a recent discussion see \cite{Gaiotto:2007qi}.}

The work of Bagger and Lambert \cite{Bagger:2007jr} presents an
explicit action and supersymmetry transformations for an ${\cal N}
=8$ Chern--Simons theory in three dimensions evading the ${\cal N}
\le 3$ bound mentioned above. Their construction can be described in
terms of an interesting new type of algebra, which we call a BL
algebra.\footnote{Gustavsson, studying the same problem in
\cite{Gustavsson:2007vu}, was independently led to formulate
conditions that are equivalent to BL algebras. The equivalence is
described in \cite{Bagger:2007vi}.} It involves a totally
antisymmetric triple bracket analog of the Lie bracket\footnote{Such brackets,
regarded as generalizations of Poisson brackets,
were considered by Nambu in 1973 \cite{Nambu:1973qe}.
For a recent discussion of Nambu brackets see \cite{Curtright:2002fd}.}
$$
[T^a, T^b, T^c] = f^{abc}{}_d T^d .
$$
There should also be a symmetric invertible metric $h^{ab}$ that can
be used to raise and lower indices. The structure constants
$f^{abcd}$ defined in this way are required to have total
antisymmetry. Furthermore, this tensor is also required to satisfy a
quadratic constraint, analogous to the Jacobi identity, which
BL call the ``fundamental equation.''

An important question, of course, is whether BL algebras have any
nontrivial realizations. BL settle this question by noting that a
solution is provided by a set of four generators $T^a$ that
transform as a four-vector of an $SO(4)$ gauge group. In this
example $f^{abcd} = \e^{abcd}$ and $h^{ab} = \d^{ab}$.
After reviewing the free theory in Section 2, this paper reviews the
BL $SO(4)$ theory in Section 3 making a couple of new observations
in the process. The first is an explicit verification that the
action is invariant under the conformal supersymmetries as well as
the Poincar\'e supersymmetries. Taken together, these generate the
entire $OSp(8|4)$ symmetry. The second is a careful demonstration in
Section~4 of a fact noted in \cite{Bagger:2007vi}, namely that the
theory is parity conserving. This feature, which is essential for a
dual to the M-theory solution, involves combining a spatial
reflection with an $SO(4) = SU(2) \times SU(2)$ reflection. The
latter reflection can be interpreted as interchanging the two
$SU(2)$ factors.

We also explore whether there exist BL theories for other choices of
gauge groups and matter representations. Motivated by the $SO(4)$
example, Section~5 considers parity-conserving theories with gauge
group $G \times G$ and matter fields belonging to a representation
$(R,R)$, where $R$ is some representation of $G$. Two classes of
such examples that have been examined carefully are based on $G=
SO(n)$ and $G = USp(2n)$ with $R$ chosen to be the fundamental
representation in each case. The first of these two classes is described
in detail. The free theory (appropriate for a single M2-brane)
appears in this classification as $G=SO(1)$, and the $SO(4)$ theory
appears as $G= USp(2)$. An invariant totally antisymmetric
fourth-rank tensor $f^{abcd}$, where $a,b,c,d$ label components of
the representation $(R,R)$, can be constructed. However, it turns
out that the fundamental equation is satisfied only for the free
theory, the $SO(4)$ theory, and the $G = SO(2)$ case. The $SO(2)$ case
does not give a new theory, however, for reasons that are explained
in the text.

BL suggested that there may be other theories with $OSp(8|4)$
superconformal symmetry based on nonassociative algebras. Following up
on this suggestion, Section~5 attempts to utilize the algebra of octonions
in this manner. This leads to a seven-dimensional BL-type algebra.
However, once again it turns out that the fundamental identity is
not satisfied. Thus, this approach also does
not lead to other consistent field theories with $OSp(8|4)$
superconformal symmetry. Based on these studies, we conjecture that
the $SO(4)$ BL theory is the only nontrivial three-dimensional
Lagrangian theory with $OSp(8|4)$ superconformal symmetry, at least
if one assumes irreducibility and a finite number of fields.

It is a curious coincidence that three-dimensional gravity
with a negative cosmological constant can be formulated as a twisted
Chern--Simons theory based on the gauge group $SO(2,2)$. Aside from
the noncompact form of the gauge group, this is identical to the Chern--Simons
term that is picked out by the BL theory. This is discussed in Section~6.

\section{The Free Theory}

Let us start with the well-known free ${\cal N} =8$ superconformal
theory. It contains no gauge fields, so it is not a Chern--Simons
theory. The action is
\begin{equation} \label{free3d}
S = \half \int \left( -\pa^\m \phi^I \pa_\m \phi^I + i\ov{\psi}^A
\g^\m\pa_\m \psi^A\right)d^3 x.
\end{equation}
This theory has $OSp(8|4)$ superconformal symmetry. The R-symmetry
is $Spin(8)$ and the conformal symmetry is $Sp(4) = Spin (3,2)$. The
index $I$ labels components of the fundamental $8_v$ representation
of $Spin(8)$ and the index $A$ labels components of the spinor $8_s$
representation. In particular, $\psi^A$ denotes 8 two-component
Majorana spinors. The Poincar\'e and conformal supersymmetries
belong to the other spinor representation, $8_c$, whose components
are labeled by dotted indices $\dot A$, etc.

The three inequivalent eight-dimensional representations of
$Spin(8)$ can couple to form a singlet. The invariant tensor (or
Clebsch--Gordan coefficients) describing this is denoted $\G^I_{A
\dot A}$, since it can be interpreted as eight matrices satisfying a
Dirac algebra. We also use the transpose matrix, which is written
$\G^I_{\dot A A}$ without adding an extra symbol indicating that it
is the transpose. These matrices have appeared many times before in
superstring theory.

Note that in our conventions $\g^\m$ are $2\times2$ matrices and
$\G^I$ are $8 \times 8$ matrices. They act on different vector
spaces and therefore they trivially commute with one another.
BL use a somewhat different formalism in which $\g^\m$ and
$\G^I$ are 11 anticommuting $32 \times 32$ matrices. We find this
formalism somewhat confusing, since the three-dimensional theories
in question cannot be obtained by dimensional reduction of a
higher-dimensional theory (in contrast to ${\cal N} =4$ super
Yang--Mills theory).

The action (\ref{free3d}) is invariant under the supersymmetry
transformations
\begin{equation}
\d \phi^I = i\ov\e^{\dot A} \G^I_{\dot A A} \psi^A = i\ov\e \G^I
\psi = i\ov\psi \G^I \e
\end{equation}
\begin{equation}
\d \psi =  -\g \cdot \pa \phi^I \G^I \e.
\end{equation}
One can deduce the conserved supercurrent by the Noether method,
which involves varying the action while allowing $\e$ to have
arbitrary $x$ dependence. This gives
\[
\d S = - i \int \pa_\m \ov\e \G^I \g\cdot\pa\phi^I \g^\m \psi d^3 x.
\]
Thus the conserved supercurrent is $i \G^I \g\cdot\pa\phi^I \g^\m
\psi$. The conservation of this current is easy to verify using the
equations of motion.

Let us now explore the superconformal symmetry. As a first try, let
us consider taking $\e^{\dot A}(x) = \g\cdot x \eta^{\dot A}$, since
this has the correct dimensions. Using $\pa_\m\e(x) = \g_\m \eta$
and $\g^\m \g^\rho \g_\m = - \g^\rho$, this gives
\[
\d S =  i\int \ov\psi \g\cdot \pa \phi^I \G^I \eta d^3x.
\]
This can be canceled by including an additional variation of the
form $\d\psi \sim \G^I \phi^I \eta$. Thus the superconformal
symmetry is given by
\begin{equation}
\d \phi^I  = i\ov\psi \G^I \g\cdot x \eta
\end{equation}
\begin{equation}
\d \psi =  -\g \cdot \pa \phi^I \G^I \g \cdot x \eta - \phi^I \G^I
\eta.
\end{equation}
One can deduce the various bosonic $OSp(8|4)$ symmetry
transformations by commuting $\e$ and $\eta$ transformations. Of
these only the conformal transformation, obtained as the commutator
of two $\eta$ transformations, is not a manifest symmetry of the
action. It is often true that scale invariance implies conformal
symmetry. However, this is not a general theorem, so it is a good
idea to check conformal symmetry explicitly as we have done.

\section{The $SO(4)$ theory}

The $SO(4)$ gauge theory contains scalar fields $\phi^I_a$ and
Majorana spinor fields $\psi^A_a$ each of which transform as
four-vectors of the gauge group ($a =1,2,3,4$). In addition there
are $SO(4)$ gauge fields $A^{ab}_\m$ with field strengths
$F^{ab}_{\m\n}$. Since four-vector indices are raised and lowered
with a Kronecker delta, we do not distinguish superscripts and
subscripts. $A$ and $F$ are called $\tilde A$ and $\tilde F$ by BL.

The action is a sum of a matter term and a Chern--Simons term:
\begin{equation}
S_k = k\left( S_{\rm m} + S_{\rm CS} \right).
\end{equation}
We choose normalizations such that the level-$k$ action $S_k$ is $k$
times the level-one action $S_1$. Then $k$, which is a positive
integer, is the only arbitrary parameter. Perturbation theory is an
expansion in  $1/k$. So the theory is weakly coupled and can be
analyzed in perturbation theory when $k$ is large. The goal here is
to construct and describe the classical action.

The required level-one Chern--Simons action is given by
\begin{equation}
S_{\rm CS} = \a\int  \tilde\om_3,
\end{equation}
where the ``twisted'' Chern--Simons form $\tilde\om_3$ is
constructed so that
\begin{equation}
d\tilde\om_3 = \half \ep_{abcd} F_{ab} \wedge F_{cd}.
\end{equation}
This implies that
\begin{equation}
\tilde\om_3 = \half \ep_{abcd} A_{ab} \wedge ( dA_{cd} + \frac{2}{3}
A_{ce} \wedge A_{ed}).
\end{equation}
When $SO(4)$ is viewed as $SU(2)\times SU(2)$, this is the
difference of the Chern--Simons terms for the two $SU(2)$ factors. The
coefficient $\a$ is chosen so that these have standard level-one
normalization. Varying the gauge field by an amount $\d A$, one has
(up to a total derivative)
$$
\d \tilde\om_3 = \ep_{abcd} \d A_{ab} \wedge F_{cd}
$$
or
$$
\d S_{\rm CS} = \frac{\a}{2}\int \ep_{abcd} \, \ep^{\m\n\r}\d A^{ab}_\m
F^{cd}_{\n\r} d^3 x.
$$

The $SO(4)$ matter action is a sum of kinetic and interaction terms
\begin{equation}
S_{\rm m} = S_{\rm kin} + S_{\rm int},
\end{equation}
where
\begin{equation}
S_{\rm kin} = \int d^3 x \left(- \half (D_\m \phi^I)_a (D^\m
\phi^I)_a + \frac{i}{2} \ov\psi_a \g^\m (D_\m \psi)_a \right)
\end{equation}
and
\begin{equation}
S_{\rm int} = \int d^3 x \left(ic \, \ep_{abcd} \ov\psi_a \G^{IJ}
\psi_b \phi^I_c \phi^J_d - \frac{4}{3}c^2  \sum (\ep_{abcd} \phi^I_b
\phi^J_c \phi^K_d )^2 \right).
\end{equation}

The supersymmetry transformations that leave the action invariant
are
\begin{equation}
\d \phi^I_a =  i\ov\e \G^I \psi_a
\end{equation}
\begin{equation}
\d \psi_a =  -\g^\m (D_\m \phi^I)_a \G^I \e +\frac{2c}{3} \ep_{abcd}
\G^{IJK}\e\phi^I_b \phi^J_c \phi^K_d
\end{equation}
\begin{equation}
\d A_{\m ab} = 4ic \, \ep_{abcd} \, \ov\psi_c\g_\m  \G^I \phi^I_d \e
\end{equation}
for the identification
$$
c = \frac{1}{16\a}.
$$
The formulas agree with BL for $c=3$, which corresponds to $\a =
1/48$. Any apparent minus-sign discrepancies are due to the
different treatment of the Dirac matrices discussed earlier.

The conformal supersymmetries also hold. They can be analyzed in the
same way that was discussed for the free theory. The result, as
before, is to replace $\e$ by $\g\cdot x \eta$ and to add a term
$-\phi^I_a \G^I\eta$ to $\d \psi_a$. We have verified the Poincar\'e
and the conformal supersymmetries of this theory in complete detail.
Thus this theory has $OSp(8|4)$ superconformal symmetry and $SO(4)$
gauge symmetry. It also has parity invariance, which we explain in
the next section.

\section{Parity Conservation}

The relative minus sign between the two $SU(2)$ contributions to the
Chern--Simons term has an interesting consequence. Normally,
Chern--Simons theories are parity violating. In this case, however,
one can define the parity transformation to be a spatial reflection
together with interchange of the two $SU(2)$ gauge groups. Then one
concludes that the Chern--Simons term is parity
conserving.\footnote{This was pointed out to us by A. Kapustin
before the appearance of \cite{Bagger:2007vi}. This way of implementing
parity conservation, including the odd parity of a spinor bilinear, was
understood already in \cite{Deser:1981wh}.}

To conclude that the entire theory is parity-conserving, there is
one other term that needs to be analyzed. It is the one that has the
structure
$$
\ep_{abcd} \bar \psi_a \G^{IJ} \psi_b \phi^I_c \phi^J_d.
$$
The interchange of the two $SU(2)$ groups gives one minus sign (due
to the epsilon symbol), so invariance will only work if a spinor
bilinear of the form $\bar\psi_1 \psi_2 = \psi_1^{\dagger} \g^0
\psi_2$ is a pseudoscalar in three dimensions. So we must decide
whether this is true. Certainly, in four dimensions such a structure
is usually considered to be a scalar. The R-symmetry labels are
irrelevant to this discussion.

Let us review the parity analysis of spinor bilinears in four
dimensions. The usual story is that the parity transform (associated
to spatial inversion $\vec{x} \to - \vec{x}$) of a spinor is given
by $\psi \to \g^0 \psi$. There are two points to be made about this.
First, spatial inversion is a reflection in four dimensions. This
differs from the case in three-dimensional spacetime, where spatial
inversion is a rotation, rather than a reflection. Therefore, it is
more convenient for generalization to the three-dimensional case to
consider a formula for the transformation of a spinor under
reflection of only one of the spatial coordinates ($x^i$, say).
Under this reflection, the formula in four dimensions is $\psi \to i
\g^i \g_5 \psi$. For this choice reflecting all three coordinates
gives the previous rule $\psi \to \g^0 \psi$ (up to an ambiguous and
irrelevant sign). With this rule, one can easily show that
$\bar\psi_1 \psi_2$ is a scalar and $\bar\psi_1 \g_5 \psi_2$ is a
pseudoscalar, as usual.

The second point is that the Dirac algebra for four-dimensional
spacetime has an automorphism $\g^\m \to i \g^\m \g_5$. In other
words,
$$
\{i \g^\m \g_5, i \g^\n \g_5\} = \{\g^\m , \g^\n \} = 2 \eta^{\m\n}.
$$
This automorphism squares to $\g^\m \to - \g^\m$, which is also an
automorphism. The kinetic term, which involves $\ov\psi \g\cdot
\pa\psi$, is invariant under this automorphism, since $i\g^0
\g_5i\g^\m \g_5= \g^0 \g^\m$. In view of this automorphism, it is
equally sensible to define a reflection by the rule $\psi \to \g^i
\psi$. However, if one makes this choice, then one discovers that
$\bar\psi_1 \psi_2$ is a pseudoscalar and $i\bar\psi_1 \g_5 \psi_2$
is a scalar. This makes sense, since they (and their negatives) are
interchanged by the automorphism.

In the case of three dimensions, there is no analog of $\g_5$, and so the
automorphism discussed above has no analog. As a result, the only sensible
rule for a reflection is $\psi \to \g^i  \psi$. Then one is forced to
conclude (independent of any conventions) that $\bar\psi_1 \psi_2$ is a
pseudoscalar. This is what we saw is required for the $SO(4)$
super Chern--Simons theory to be parity conserving.

\section{The Search for Generalizations}

Possible generalizations of the $SO(4)$ theory are suggested by the
fact that $SO(4) = SU(2) \times SU(2) = USp(2) \times USp(2)$ and
that a four-vector field $\phi^a$ can be reexpressed as a bifundamental
field  $\phi^{\a\a'}$.

An infinite class of candidate theories with the same type of
structure is based on the gauge group $SO(n) \times SO(n)$ with
matter fields $\phi^{\a\a'}$ assigned to the bifundamental
representation $({\bf n,n})$. In this case one takes the gauge field
to be
\begin{equation}
A_{\a\a'\b\b'} = \d_{\a\b} A'_{\a'\b'} + \d_{\a'\b'}A_{\a\b},
\end{equation}
where $A_{\a\b} = -A_{\b\a}$ and $A'_{\a'\b'}= - A'_{\b'\a'}$ are
$SO(n)$ gauge fields. The $n=1$ case is the free theory with 8
scalars and 8 spinors and no gauge fields, which was discussed in
Section 2.


The BL structure constants vanish for $n=1$, and for $n>1$ they are
given by
\begin{equation} \label{Ofabcd}
f^{\a\a'\b\b'\g\g'\d\d'} = \frac{1}{2(n-1)}\Big(-\d^{\a\b}\d^{\g\d}\d^{\a'\d'}\d^{\b'\g'}
+\d^{\a\b}\d^{\g\d}\d^{\a'\g'}\d^{\b'\d'}
\end{equation}
$$
-\d^{\a\g}\d^{\d\b}\d^{\a'\b'}\d^{\g'\d'}
+\d^{\a\g}\d^{\d\b}\d^{\a'\d'}\d^{\g'\b'}
-\d^{\a\d}\d^{\b\g}\d^{\a'\g'}\d^{\d'\b'}
+\d^{\a\d}\d^{\b\g}\d^{\a'\b'}\d^{\d'\g'}\Big).
$$
For this choice one finds that the dual gauge field is
$$
\tilde A^{\a\a'\b\b'} = f^{\a\a'\b\b'\g\g'\d\d'}A_{\g\g'\d\d'}
= \d^{\a\b} A'^{\a'\b'} - \d^{\a'\b'}A^{\a\b}.
$$
Therefore the twisted Chern--Simons term again is proportional to
the difference of the individual Chern--Simons terms, as required by
parity conservation. However, the BL fundamental equation is not
satisfied for $n>2$, and there are a number of inconsistencies in
the supersymmetry algebra. This leaves the $n=2$ case as the only
remaining candidate for a new theory. This theory (if it exists) has
the same matter content as the BL theory, but fewer gauge fields.
Even though the BL algebra is okay in this case, the elimination of
four gauge fields gives a violation of another requirement.
Specifically, the antisymmetric tensor $f_{abcd}$ is not $SO(2)
\times SO(2)$ adjoint valued in a pair of indices. This is an
essential requirement, because the formula for the supersymmetry
variation of the gauge field has the form
\begin{equation}
\d A_{\m ab} = 4ic \, f_{abcd} \, \ov\psi_c\g_\m  \G^I \phi^I_d \e.
\end{equation}
This equation does not make sense when the right-hand side
introduces unwanted degrees of freedom that do not belong to the
adjoint representation. This problem arises for all cases with $n>1$
including the $n=2$ case in particular. One could try to remove the
nonadjoint pieces of the right-hand side, but that leads to other
inconsistencies.

A completely analogous analysis exists for candidate theories based
on the gauge group $USp(2n) \times USp(2n)$ with matter fields
belonging to the bifundamental representation. For the choice $n=1$
this is the $SO(4)$ theory of Section 3. Again, one can construct a
totally antisymmetric tensor $f^{abcd}$ for all $n$. However, this does
give any new theories, because the BL fundamental equation is not
satisfied for $n>1$.

Let us now describe another attempt to construct new examples. BL
describe a systematic way to obtain totally antisymmetric triple
brackets based on nonassociative algebras. However, the examples
they discuss all involve adjoining ``a fixed Hermitian matrix $G$''
that does not seem to be compatible with a conventional Lie algebra
interpretation. Here we explore dispensing with such an auxiliary
matrix and applying their procedure to the most familiar
nonassociative algebra we know, namely the algebra of octonions. The
question to be addressed is then whether this gives a new
superconformal theory with the gauge group $G_2$ and with the matter
fields belonging to the seven-dimensional representation.

Let us denote the imaginary octonions by $e_a$ with
$a=1,2,\ldots,7$. These have the nonassociative multiplication table
$$
e_a e_b = t_{abc} e_c - \d_{ab}.
$$
The totally antisymmetric tensor $t_{abc}$ has the following
nonvanishing components
$$
t_{124} = t_{235}= t_{346} = t_{457} = t_{561}= t_{672} = t_{713}
=1.
$$
Note that these are related by cyclic permutation of the indices
$(a,b,c) \to (a+1,b+1,c+1)$. It is well known that $t_{abc}$ can be
regarded as an invariant tensor describing the totally antisymmetric
coupling of three seven-dimensional representations of the Lie group $G_2$.

Let $T_{ab}$ denote a generator of an $SO(7)$ rotation in the $ab$
plane. The $SO(7)$ Lie algebra is
$$
[T_{ab}, T_{cd}] = T_{ad} \d_{bc} - T_{bd} \d_{ac} - T_{ac} \d_{bd}
+ T_{bc} \d_{ad}.
$$
The generators of $G_2$ can be described as a 14-dimensional
subalgebra of this Lie algebra. A possible choice of basis is given
by
$$
X_1 = T_{24} - T_{56} \quad \and \quad Y_1 = T_{24} - T_{37}
$$
and cyclic permutations of the indices. This gives 14 generators
$X_A$ consisting of $X_a$ and $X_{a+7} = Y_a$. By representing the
generators $T_{ab}$ by seven-dimensional matrices in the usual way,
one can represent the $G_2$ generators by antisymmetrical
seven-dimensional matrices. These can then be used in the usual way
to express $G_2$ gauge fields as seven-dimensional matrices
$A_{ab}$.

The group $G_2$ is a subgroup of $SO(7)$ in which the ${\bf 7}$ of
$SO(7)$ corresponds to the ${\bf 7}$ of $G_2$. Thus, the seven-index
epsilon symbol, which is an invariant tensor of $SO(7)$, is also an
invariant tensor of $G_2$. It can be used to derive an antisymmetric
fourth-rank tensor of $G_2$:
$$
f_{abcd} = \frac{1}{6} \ep_{abcdefg} t_{efg}.
$$
This tensor has the following nonzero components
$$
f_{7356} = f_{1467} = f_{2571} = f_{3612} = f_{4723} = f_{5134} =
f_{6245} =1.
$$
These are also related by cyclic permutations. This tensor is the same
(up to normalization) as the one given by the construction based on
associators that was proposed by BL.

If one defines
$$
[abc,def] = \sum_x f_{abcx} f_{defx},
$$
the BL fundamental equation takes the form
$$
[abw,xyz] - [abx, yzw] + [aby, zwx] - [abz, wxy] =0.
$$
Note that the left-hand side has antisymmetry in the pair $(a,b)$
and total antisymmetry in the four indices $(w,x,y,z)$. One can
verify explicitly that these relations are {\it not} satisfied by
the tensor $f_{abcd}$ given above. (BL did not claim that it
necessarily would satisfy the fundamental equation.) Thus, the
tensor $f_{abcd}$ does not define a seven-dimensional BL algebra,
and we do not obtain a new theory for the gauge group $G_2$.

\section{Relation to anti de Sitter gravity?}

Pure three-dimensional gravity with a negative cosmological constant
can be formulated as a twisted Chern--Simons theory based on the
gauge group $SO(2,2)$. \cite{Townsend,Witten:1988hc,Witten:2007kt}
The BL theory, on the other hand, requires a twisted Chern--Simons
term for the gauge group $SO(4)$. Aside from the signature, these
are exactly the same! What should one make of this
coincidence?\footnote{This section was motivated by a question
raised by Aaron Bergman at a seminar given by JHS.}

The BL theory was motivated by the desire to construct conformal
field theories dual to gravity in four-dimensional anti de Sitter
space. So the notion that it might be possible to interpret it as a
gravity theory in three-dimensional anti de Sitter space is
certainly bizarre. The BL theory can be modified easily to the gauge
group $SO(2,2)$, though this introduces some disturbing minus signs
into half of the kinetic terms of the scalar and spinor fields. If
one makes this change anyway, the Chern--Simons term is exactly that
for gravity. However, there is a serious problem with a
gravitational interpretation in addition to the problem of the
negative kinetic terms: a gravity theory should have diffeomorphism
symmetry. The Chern--Simons term has this symmetry, but the matter
terms in the Lagrangian contain the three-dimensional Lorentz metric
to contract indices, so they are not diffeomorphism invariant. Thus,
we believe that there is no sensible interpretation of the BL theory
as a three-dimensional gravity theory. Nonetheless, it is striking
that its Chern--Simons term is so closely related to the one that
arises in the Chern--Simons description of three-dimensional gravity
with a negative cosmological constant.

The $SO(2,2)$ Chern--Simons formulation of three-dimensional gravity
in anti de Sitter space has supergravity generalizations, which can
be formulated as Chern--Simons theories for the supergroups
\cite{Townsend}
$$
OSp(p|2) \times OSp (q|2).
$$
The pure gravity case corresponds to $p=q=0$.
The existence of these supergravity theories, together with
the bizarre coincidence noted above, suggests
trying to generalize the BL theory to the corresponding supergroup
extensions of $SO(4)$. This idea encounters problems with spin and
statistics, since the odd generators of this supergroup are not
spacetime spinors.

\section{Conclusion}

We have studied classical Lagrangian theories in three dimensions
with $OSp(8|4)$ superconformal symmetry. This symmetry and parity
conservation were explicitly verified for the free theory and the
Bagger--Lambert $SO(4)$ theory. A search for further examples of
such theories was described. This work led us to conjecture that
there are no other such theories, at least if one assumes a finite
number of fields.

The relevance of these superconformal Chern--Simons theories to
AdS/CFT is an intriguing question. The free theory (associated to a
single M2-brane) is presumably dual to the $AdS_4 \times S^7$
solution with one unit of flux. Based on an analysis of the moduli
space of classical vacua, BL proposed in \cite{Bagger:2007vi} that
the $SO(4)$ theory is dual to $AdS_4 \times S^7$ with three units
of flux, but they do not discuss how to choose the level $k$.

To conclude, maximally supersymmetric conformal field theories with
a Lagrangian formulation are not common. The BL theory is the first
nontrivial example (above two dimensions) since the construction of
${\cal N}=4$ super Yang--Mills theory over 30 years ago. Thus, we
expect that this theory will play a role in the future
development of string theory and M-theory, but it is unclear to us
what that role will be.

\section*{Acknowledgments}

We are grateful to Jonathan Bagger for bringing \cite{Bagger:2007jr}
to our attention. We have benefitted from discussions with Anton
Kapustin and from Aaron Bergman's question about the relationship to
three-dimensional gravity. This work was supported in part
by the U.S. Dept. of Energy under Grant No. DE-FG03-92-ER40701.

\end{document}